\documentclass[twocolumn,amsmath,amssymb,prc,superscriptaddress,floatfix,showpacs,nofootinbib]{revtex4}
\usepackage[dvips]{epsfig}
\usepackage{slashed}
\usepackage{mathrsfs}
\usepackage{graphicx}
\usepackage{xcolor}
\usepackage{hyperref}
\usepackage{dcolumn}
\usepackage{subfigure}
\usepackage{xspace}
\usepackage{bbm}

\graphicspath{
{./}
{./Figures/}
}
\begin{document}
\title{Universality far from equilibrium: From superfluid Bose gases to heavy-ion collisions}

\author{J.~Berges}
\affiliation{Institut f\"{u}r Theoretische Physik, Universit\"{a}t Heidelberg,
  Philosophenweg 16, 69120 Heidelberg, Germany}
\affiliation{ExtreMe Matter Institute EMMI, GSI Helmholtzzentrum,
  Planckstra\ss e 1, 64291 Darmstadt, Germany}

\author{K.~Boguslavski}
\affiliation{Institut f\"{u}r Theoretische Physik, Universit\"{a}t Heidelberg,
  Philosophenweg 16, 69120 Heidelberg, Germany}

\author{S.~Schlichting}
\affiliation{Brookhaven National Laboratory, Physics Department, Bldg. 510A, Upton, NY 11973, USA}

\author{R.~Venugopalan}
\affiliation{Brookhaven National Laboratory, Physics Department, Bldg. 510A, Upton, NY 11973, USA}

\begin{abstract}
Isolated quantum systems in extreme conditions can exhibit unusually large occupancies per mode. This over-population gives rise to new universality classes of many-body systems far from equilibrium. We present theoretical evidence that important aspects of non-Abelian plasmas in the ultra-relativistic limit admit a dual description in terms of a Bose condensed scalar field theory.
\end{abstract}
\maketitle


\section{Introduction}

In recent years there have been important advances in understanding isolated quantum systems in extreme conditions far from equilibrium. Prominent examples include the (pre-)heating process in the early universe after inflation, the initial stages in collisions of ultra-relativistic nuclei at giant laboratory facilities, as well as table-top experiments with ultracold quantum gases. Even though the typical energy scales of these systems vastly differ, they can show very similar dynamical properties. Certain characteristic numbers can even be quantitatively the same. One may use this universality to learn from experiments with cold atoms aspects about the dynamics during the early stages of our universe~\cite{Schmiedmayer:2013xsa}.

Ultracold quantum gases are known to exhibit universal properties near unitarity in the presence of a very large scattering length $a$~\cite{Braaten:2004rn}.\footnote{Here three spatial dimensions are considered and natural units will be employed where the reduced Planck constant ($\hbar$), the speed of light ($c$) and Boltzmann's constant ($k_B$) are set to one.} 
In this work we consider a different universal regime away from unitarity, which occurs far from equilibrium and has attracted much interest recently in the context of nonthermal fixed points~\cite{Berges:2008wm,Berges:2008sr,Scheppach:2009wu}. For an interacting Bose gas of density $n$ with an inverse coherence length described by the momentum $Q = \sqrt{16\pi a n}$, this novel regime is characterized by an unusually large mode occupancy $f(Q) \sim 1/\zeta $ in the dilute regime where $\zeta = \sqrt{n a^3} \ll 1$. The average density $n = \int d^3p/(2\pi)^3 f(p) \sim Q^3/\zeta$ becomes parametrically large, reflecting the underlying nonequilibrium distribution $f(p)$ of modes. Because of the large typical occupancies, the system is strongly correlated and its properties become insensitive to the details of the underlying model or initial conditions. The far-from-equilibrium behavior can be described in terms of universal exponents and scaling functions~\cite{Berges:2008wm,Berges:2008sr,Scheppach:2009wu,Nowak:2010tm,Berges:2012us,Nowak:2012gd}, similar to the description of critical phenomena in thermal equilibrium. 

Such an over-occupation of modes can be found in a variety of systems in extreme conditions. In heavy-ion collisions at ultra-relativistic energies, a nonequilibrium plasma of highly occupied gluon fields with characteristic momentum $Q_s$ is expected to form shortly after the collision~\cite{Gelis:2010nm,Lappi:2006fp}. While the running gauge coupling $\alpha(Q_s)$ is weak for sufficiently large $Q_s$, the system becomes strongly correlated because the typical gluon occupancy $f_g(Q_s) \sim 1/\alpha$ is large. Here the coupling plays the corresponding role to the diluteness parameter $\zeta$. Indeed, scaling behavior has been observed in simulations of the space-time evolution of non-Abelian plasmas~\cite{Berges:2008mr,Schlichting:2012es,Kurkela:2012hp,Berges:2013eia,Kurkela:2014tea}. 

In this letter we present first theoretical evidence that important universal properties of these very different systems far from equilibrium can agree. Since symmetries and underlying scattering processes for scalar and gauge theories show profound differences, the observation of universal dynamics in this case is highly non-trivial. We employ the largest real-time lattice gauge theory simulations to date to analyze the space-time evolution of non-Abelian plasmas in the limit of ultra-relativistic energies~\cite{Berges:2013eia}. 
These results are compared to the nonequilibrium dynamics of scalar Bose fields. To put our studies in the context of heavy-ion collision experiments, in both cases we investigate longitudinally expanding systems in three space dimensions. We consider a relativistic bosonic field theory with weak coupling parameter $\lambda$, for which suitable atomic model Hamiltonians may be constructed~\cite{Endres:2012wg}.
We will see below that characteristic low-momentum properties turn out to be in the same universality class as the corresponding non-relativistic theory. Moreover, we generalize our bosonic system to include $N$ real-valued field components to study the possible dependence of the universality class on $N$. For the non-Abelian gluon fields we consider two different colors, since no indications for a significant dependence on a larger number of colors have been reported so far.   

\section{Dynamic universality class}

\begin{figure}[t]						
 \centering
 \includegraphics[width=0.5\textwidth]{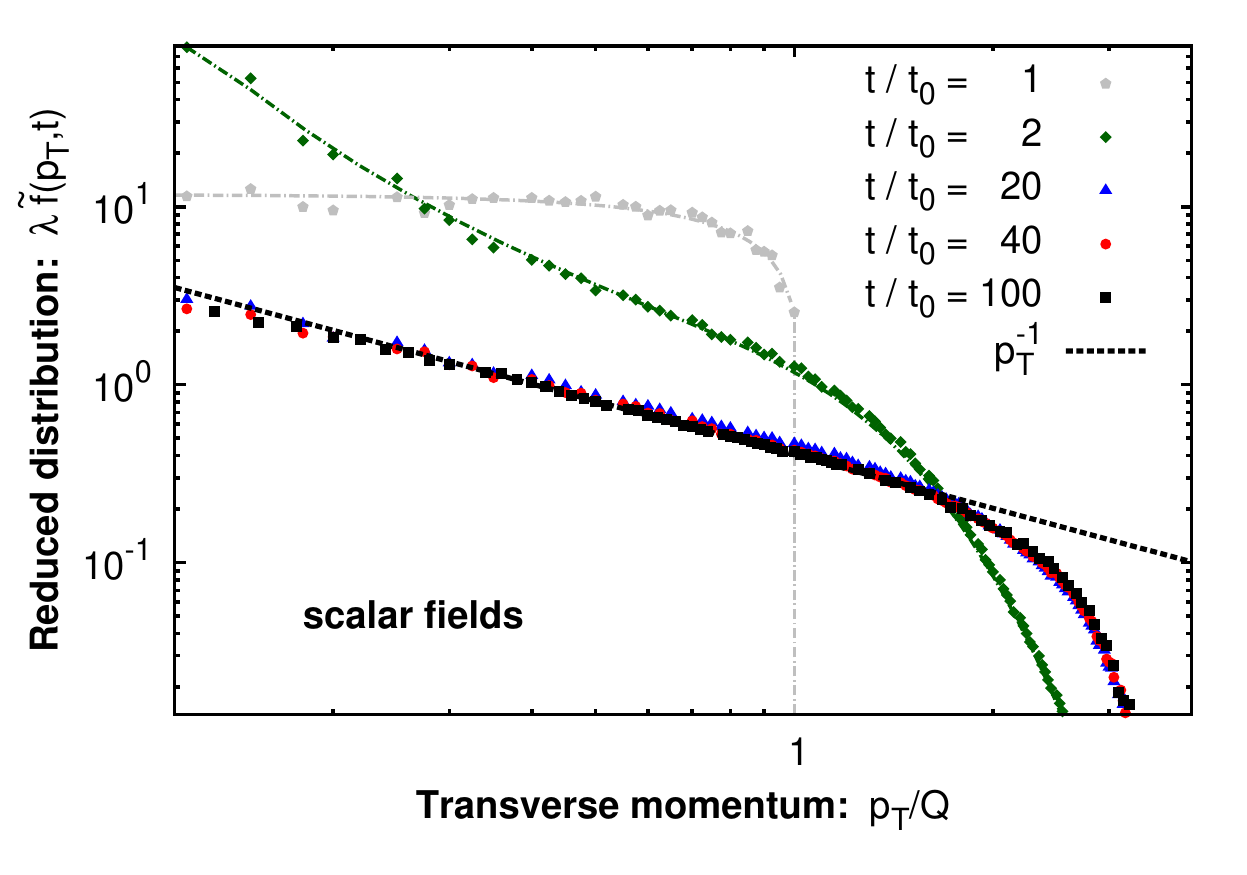}
  \caption{The scalar reduced distribution quickly approaches a time-independent form, with $\lambda \tilde{f}(p_T,t) \sim 1/p_T$ for $p_T \lesssim Q$.}
  \label{fig:intdnu}
\end{figure}

Universal scaling behavior of many-body systems in vacuum or thermal equilibrium are known to be efficiently classified in terms of universality classes, which are characterized by scaling exponents and scaling functions. Depending on the range of momenta, in equilibrium one distinguishes between infrared and ultraviolet fixed points and associated scaling properties. In general, scaling behavior of systems far from equilibrium can exhibit several distinct momentum regimes. Related examples for inertial ranges of momenta showing scaling behavior have been given in the context of wave turbulence~\cite{Micha:2004bv,Berges:2008wm,Berges:2008sr,Scheppach:2009wu,Nowak:2010tm,Berges:2012us,Nowak:2012gd,Berges:2013lsa}.

For the following presentation, we classify dynamic universality classes in terms of scaling properties of a time-dependent distribution function and refer for further discussions in terms of nonthermal renormalization group fixed points to the literature (see e.g.~\cite{Berges:2008sr}). For longitudinally expanding systems the distribution function depends on proper time $t$, on transverse momentum $p_T$, and on longitudinal momentum $p_z = \nu/t$ where $\nu$ denotes the rapidity wave number.\footnote{The space-time $(x^0,\boldsymbol{x})$ evolution with expansion along $z=x^3$ is described in terms of proper time $t=\sqrt{(x^0)^2 - (x^3)^2}$ and rapidity $\eta={\rm atanh}(x^3/x^0)$ with corresponding rapidity wave number $\nu$. 
Transverse coordinates are denoted by $x_T=(x^1,x^2)$.} In the universal regime the distribution is then determined by a time-{\it independent} scaling function $f_S$, an overall scaling with time described by the scaling exponent $\alpha$, and two exponents $\beta$ and $\gamma$ for the scaling with transverse and longitudinal momentum:
\begin{equation}
f(p_T,p_z,t) = (Q t)^\alpha f_S \left((Q t)^\beta p_T, (Q t)^{\gamma} p_z \right) \, . 
\label{eq:scaling}
\end{equation}
Systems belonging to the same universality class have the same values for $\alpha$, $\beta$ and $\gamma$ as well as the same form of $f_S(p_T,p_z)$ in a given inertial range of momenta.

\begin{figure}[t]						
 \centering
 \includegraphics[width=0.5\textwidth]{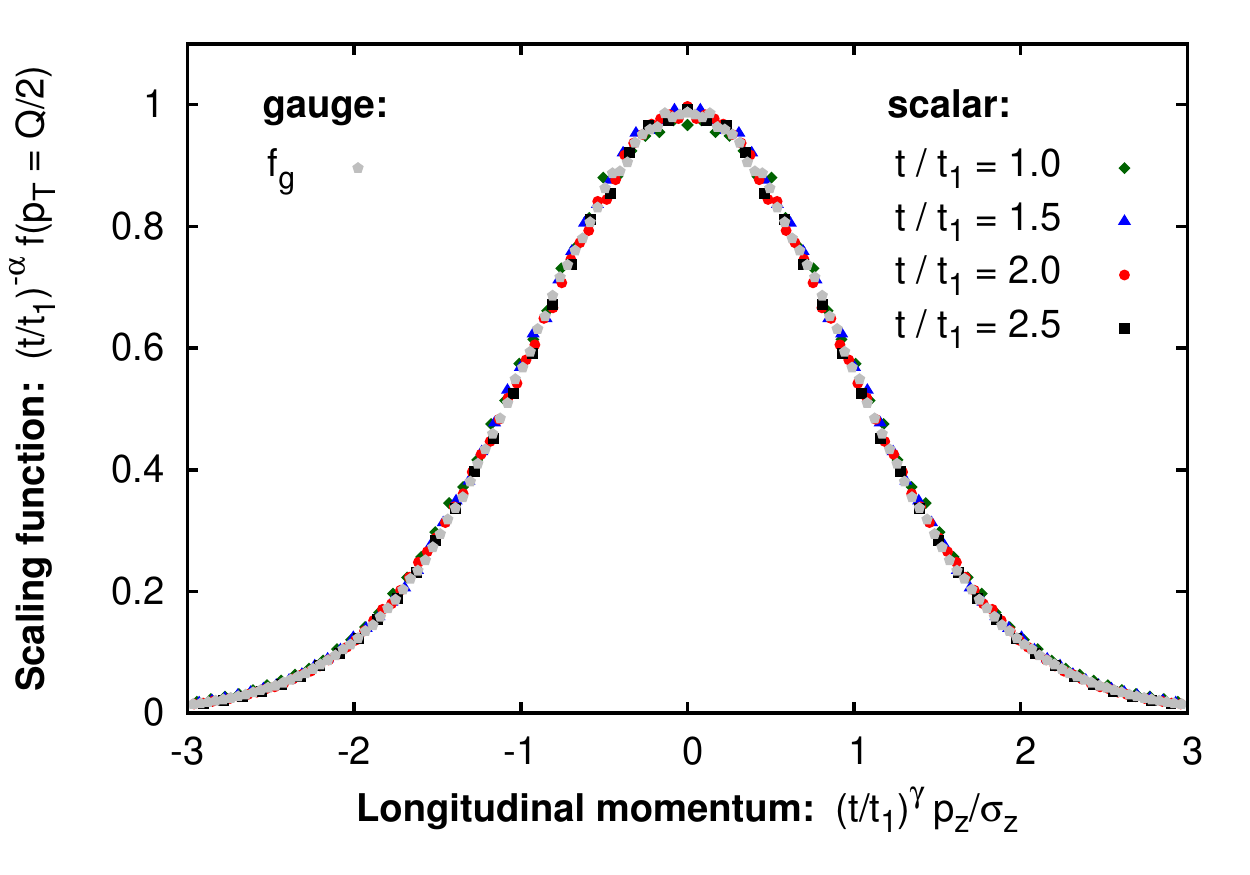}
  \caption{The normalized fixed point distribution ($t_1 = 40 t_0$) for the scalar theory compared to the gauge theory.}
  \label{fig:longSpecMid}
\end{figure}

The distribution function reflects properties of equal-time correlation functions of the underlying quantum field theory. For the massless scalar theory with quartic interaction $\lambda (\sum_a \Phi_a\Phi_a)^2/4!N$ for the $a=1,\ldots,N$ field components, the anti-commutator expectation value $F (\boldsymbol{x},\boldsymbol{x}';t,t')= \sum_a \langle \{\Phi_a(\boldsymbol{x},t),\Phi_a(\boldsymbol{x}',t')\}\rangle/2N$ determines the distribution. In spatial Fourier space we define $\ddot{F}(\boldsymbol{p};t)\equiv t t' \partial_t\partial_{t'}F(\boldsymbol{p};t,t')|_{t=t'}$ and $\dot{F}(\boldsymbol{p};t)\equiv (t \partial_t F(\boldsymbol{p};t,t') + t' \partial_{t'} F(\boldsymbol{p};t,t'))|_{t=t'}/2$, where time factors are due to the expansion. The distribution function reads $f(\boldsymbol{p},t) + 1/2 = \sqrt{F(\boldsymbol{p};t) \ddot{F}(\boldsymbol{p};t) - \dot{F}(\boldsymbol{p};t)^2}$~\cite{Berges:2008wm,Berges:2013lsa}. A similar definition can also be given for the gauge field theory with additional (Coulomb) gauge fixing~\cite{Berges:2013eia}. 

We will also assume that the quantum field theory and the corresponding classical-statistical field theory are in the same universality class for sufficiently high occupancies, and employ standard lattice simulation techniques to compute the nonequilibrium time evolution~\cite{Berges:2013eia}. The agreement is well established for critical phenomena in thermal equilibrium and has been verified explicitly for the nonequilibrium scalar field theory in the highly occupied regime, where $f(\boldsymbol{p},t) \gg 1/2$ such that the `quantum-half' becomes insignificant~\cite{Berges:2008wm}.   

We start our evolution at time $t_0$ from over-populated initial conditions, with distribution function $f(p_T,p_z,t_0) = (n_0/\lambda) \Theta ( Q - \sqrt{p_T^2+(\xi_0 p_z)^2} )$, where $n_0$ parametrizes the initial amplitude and $\xi_0$ the initial anisotropy of the distribution function. Universal results are independent of the choice of the initial conditions and the coupling, and we have varied $t_0$, $n_0$, $\xi_0$ and $\lambda$ to verify this. If not stated otherwise, we present results of the scalar theory for $n_0 = 35$, $\xi_0 = 1$ and $Q t_0 = 10^3$.\footnote{For the scalar theory we employ up to $192^2 \times 768$ lattices with spacings $Qa_T = 0.5 - 1.1$ and $Qt_0 a_\eta = 0.04 - 0.075$ in transverse and longitudinal directions. For the gauge theory, we use $256^2 \times 2048$ lattices and refer to~\cite{Berges:2013eia} for different parameter sets.}

\begin{figure}[t]						
 \centering
 \includegraphics[width=0.5\textwidth]{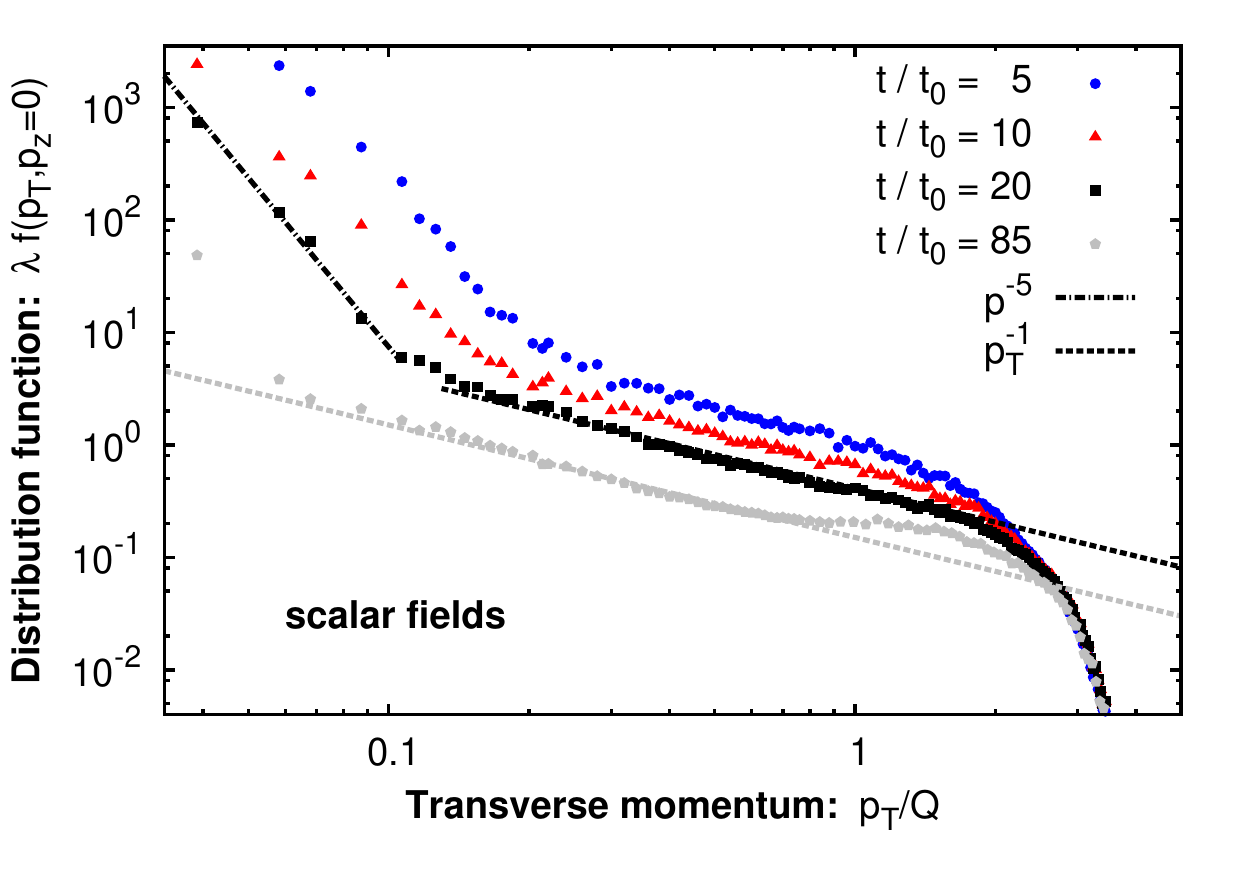}
  \caption{The distribution function for the scalar theory at $p_z=0$ for different times up to $t/t_0 = 85$.}
  \label{fig:pTdistribution}
\end{figure}
    
Fig.~\ref{fig:intdnu} shows scalar field theory results for the reduced distribution
\begin{eqnarray}
 \tilde{f}(p_T,t) = \frac{t}{t_0} \int \frac{d p_z}{2\pi Q} f(p_T,p_z,t)
\end{eqnarray}
integrated over rapidity wave number $\nu = t p_z$. It is plotted as a function of transverse momentum for $N=4$ at different times. One observes that it quickly becomes approximately time-independent.
With (\ref{eq:scaling}) follows $\tilde{f}(p_T,t) \sim t^{\alpha - \gamma +1} \int dp_z\, f_S( (Qt)^{\beta}  p_T, p_z)$ such that time independence implies the scaling relations 
\begin{equation}
 \alpha - \gamma + 1 = 0 \quad , \quad
 \beta = 0\; .
 \label{eq:a-g-b-relations}
\end{equation}
Remarkably, the very same relations have been observed in Ref.~\cite{Berges:2013eia} for the non-Abelian gauge theory. We have explicitly verified that the same relations hold in the scalar field theory case also for $N=2$.

In order to clarify the physics that determines the universality class, we note that Fig.~\ref{fig:intdnu} also exhibits a rather accurate power law $\tilde{f}(p_T,t) \sim 1/p_T$ for $p_T \lesssim Q$. Hence the particle number per transverse momentum mode $t\,dn/dp_T \sim p_T \tilde{f}$ is uniformly distributed over transverse momenta and constant as a function of time. Since the longitudinal momenta are red-shifted because of expansion, one finds $p_z \ll p_T$ for typical momentum modes. Accordingly, the energy distribution per transverse mode $t\,d\varepsilon/dp_T \sim p_T^2 \tilde{f}$ is also independent of time. Thus there is effectively neither a particle nor an energy flux in transverse momentum, which is implied by $\beta=0$ in (\ref{eq:a-g-b-relations}).

The existence of a scaling solution where both energy and particle number are conserved locally in momentum space is remarkable when contrasted to the discussion in isotropic systems without longitudinal expansion~\cite{Micha:2004bv}. There is no single scaling solution conserving both energy and particle number in the latter case. Instead, a dual cascade emerges such that in a given momentum range only one conservation law constrains the scaling
solution~\cite{waveturbulence,Berges:2008wm,Nowak:2010tm}. 
The fact that for longitudinally expanding systems both conservation laws are effective in the same momentum regime works in favor of a large universality class encompassing rather different systems. For instance, the exponent $\beta = 0$ is entirely fixed by enforcing both conservation laws without further knowledge about the underlying dynamics. Consequently, one may expect the same exponent to appear in a larger class of isolated systems, which are dominated by number conserving processes and undergo a longitudinal expansion. 
In contrast, $\alpha$, $\gamma$ and the scaling function $f_S$ are not fixed by conservation laws and the observation that theories with different symmetry groups and number of field components can exhibit common universal aspects is intriguing. 

\begin{figure}[t]						
 \centering
 \includegraphics[width=0.5\textwidth]{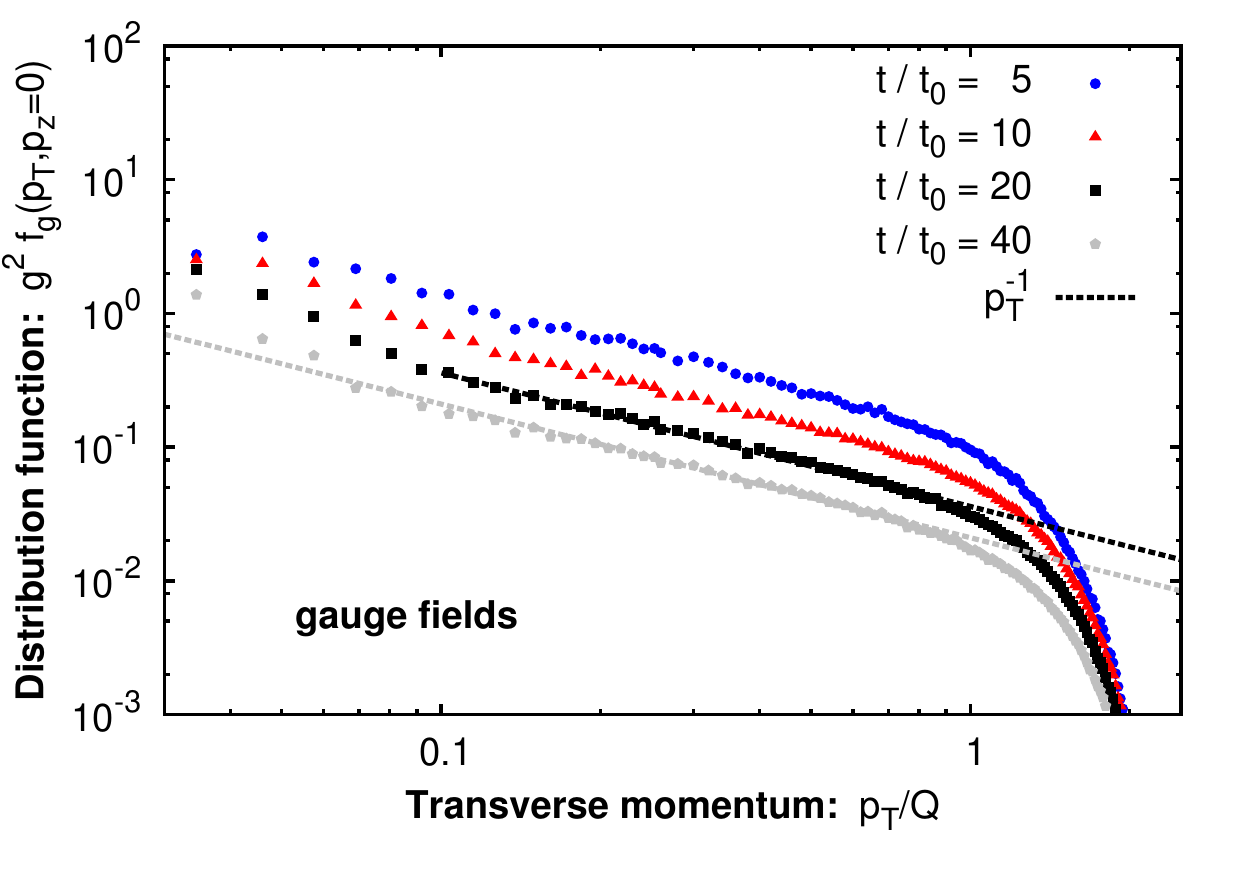}
  \caption{The gauge theory distribution function at $p_z=0$ for different times up to $t/t_0 = 40$, with $g^2 \equiv 4 \pi \alpha$.}
  \label{fig:ptgauge}
\end{figure}

The conservation laws indicate that modes are redistributed along the longitudinal direction. This has been analyzed in detail for the corresponding gauge theory in Ref.~\cite{Berges:2013eia}. Fig.~\ref{fig:longSpecMid} shows our results for the scalar theory for intermediate $p_T \sim Q/2$, where the rescaled distribution as a function of the rescaled longitudinal momentum is given for different times. According to (\ref{eq:scaling}) this fixed point distribution should be independent of time, and indeed we find that all data collapses onto a single curve using the scaling exponents
\begin{equation}
\alpha = - 2/3 \quad , \quad \gamma = 1/3 \, 
\label{eq:ab}
\end{equation}
to few percent accuracy in accordance with the scaling relation (\ref{eq:a-g-b-relations}). The very same exponents (\ref{eq:ab}) were found to characterize the gauge theory~\cite{Baier:2000sb,Berges:2013eia}. For the latter we also give our results for the normalized fixed point distribution in Fig.~\ref{fig:longSpecMid}. The Gaussian shaped curve with width squared $\sigma_z^2(p_T) = \int dp_z p_z^2 f(p_T,p_z,t_1)/\int dp_z f(p_T,p_z,t_1)$ for the scalar theory is seen to accurately agree with the corresponding gauge theory case. These results are therefore a striking manifestation of universality.    
 
\section{Inertial range and Bose condensation}

\begin{figure}[t]						
 \centering
 \includegraphics[width=0.5\textwidth]{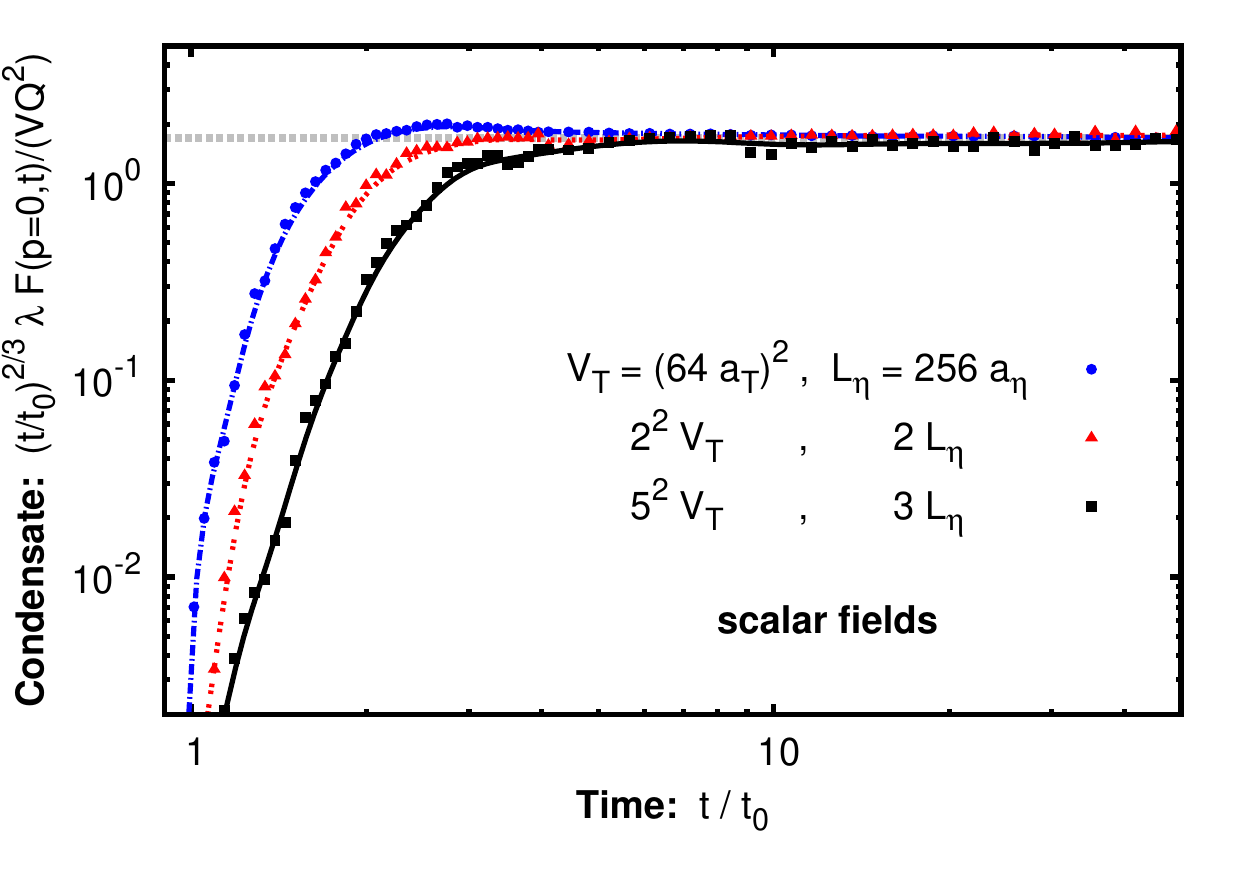}
  \caption{The scalar zero mode divided by volume $V=V_T \times L_\eta$ versus time for different volumes. (The lattice spacings for the smallest volume are $Q a_T = 0.5$ and $a_\eta = 5 \cdot 10^{-5}$.)}
  \label{fig:condensation}
\end{figure}

We shall now analyze the full momentum dependence of the spectra to determine the inertial range of this universal regime and to search for possible additional infrared and ultraviolet fixed points. 
Fig.~\ref{fig:pTdistribution} shows the scalar distribution function at vanishing longitudinal momentum for different times. The data is for $N=4$ but we find 
similar results also for $N=2$. Apart from 
$\sim 1/p_T$ behavior at intermediate momenta,
one observes a distinct infrared regime below the momentum scale where the occupation number becomes larger than $\sim 1/\lambda$. The infrared spectrum 
is {\it isotropic} in this regime~\cite{BBSV_in-prep} and described by an approximate power law $\sim 1/p^5$. We conclude that the very high occupancies at low-momenta enhance scattering rates to overcome the rate of longitudinal expansion. 
The situation is very similar to non-expanding systems, where a nonthermal infrared fixed point was observed previously for scalar field theories~\cite{Berges:2008wm,Berges:2008sr,Scheppach:2009wu,Nowak:2010tm,Berges:2012us,Nowak:2012gd,Berges:2013lsa}.
In particular, the scaling behavior agrees with the inverse particle cascade found in non-relativistic Bose gases without expansion, where the observed scaling exponent characterizes superfluid turbulence in three spatial dimensions~\cite{Scheppach:2009wu,Nowak:2010tm,Berges:2012us,Nowak:2012gd}.  

While this infrared regime limits from below the $1/p_T$ behavior at intermediate momenta, Fig.~\ref{fig:pTdistribution} shows that the bound shifts towards lower momenta at later times. The inertial range satisfying the $1/p_T$ power law increases with time on a logarithmic scale, confirming it to be a robust property of the far from equilibrium scalar dynamics. We have verified that the scaling behavior (\ref{eq:scaling}) with the exponents (\ref{eq:a-g-b-relations}) and (\ref{eq:ab}) is realized in the entire momentum region described by the $1/p_T$ power law for both the scalar and the gauge field theory. In Fig.~\ref{fig:ptgauge}, we show the corresponding results for the gauge theory distribution function.
The $\sim 1/p_T$ behavior is seen in nearly the entire momentum range. A relative enhancement is observed in the deep infrared, which is however far weaker than the corresponding behavior in the scalar theory; it remains to be seen whether this enhancement becomes stronger at later times. Regardless, 
the concept of a gauge dependent number distribution is 
problematic at low momenta and further studies in this regime should concentrate on gauge invariant correlation functions. Such an analysis would also be important in view of speculations about a strong infrared enhancement for over-populated gauge fields~\cite{Blaizot:2011xf}, where the formation of a Bose condensate is debated.

For non-expanding scalar theories, the nonthermal infrared fixed point catalyzes the formation of a Bose-Einstein condensate~\cite{Berges:2012us,Nowak:2012gd,Berloff}. We repeat the corresponding analysis in the case of longitudinal expansion and search for a scalar-field zero-mode which scales with volume, namely, $F(\boldsymbol{p}=0;t) \sim V$. This function oscillates and we take the period average to illustrate its evolution. The results are shown in Fig.~\ref{fig:condensation}, where we present the time evolution of the condensate observable for different system sizes. For the initial conditions employed, there is no condensate at time $t_0$. Accordingly, as shown in Fig.~\ref{fig:condensation}, at early times the ratio $F(\boldsymbol{p}=0;t)/(V Q^2)$ decreases as the volume is increased. However after a transient regime, the ratio becomes volume independent signaling the formation of a coherent zero mode over the entire volume.  This time-dependent condensate evolves in time as $\sim (t/t_0)^{-1/3}$. 

We turn finally to higher transverse momenta. For the scalar theory one observes from Fig.~\ref{fig:pTdistribution} that at the latest available times of $t/t_0 \sim 85$ a flat distribution for $p_T \gtrsim Q$ emerges. The distribution function in this inertial range still exhibits self-similar behavior (\ref{eq:scaling}) with the relations (\ref{eq:a-g-b-relations}) albeit with a different scaling function $f_S$. Moreover, we infer a scaling of $p_z$ characterized by the exponent $\gamma = 1/2$ in this large $p_T$ region. As a consequence, for all times considered, a significant broadening of the longitudinal distribution occurs for these hard transverse momenta as well. The systematics of this regime of hard transverse momenta will be discussed in more detail elsewhere~\cite{BBSV_in-prep}. In contrast, the gauge theory, for the shorter times explored, shows no such additional scaling regime at hard momenta.


\section{Conclusions}

We reported evidence of a novel dynamical universality class encompassing longitudinally expanding scalar fields and gauge fields over a characteristic inertial range of transverse momenta. This observed universality challenges our understanding of the thermalization process of the Quark Gluon Plasma in the limit of very high energies, where the gauge coupling is weak. Since the underlying perturbative scattering processes are very different for gauge field and scalar degrees of freedom, our findings point to a much more general principle. Here the classification of strongly correlated quantum many-body systems in terms of nonequilibrium universality classes and associated scaling properties represents a crucial step. Such a development also opens intriguing new perspectives to experimentally access universal properties of systems in extreme conditions with the help of quantum degenerate gases.\\     


\textbf{Acknowledgements:} We thank T.~Gasenzer, D.~Gelfand, V.~Kasper, A.~Pineiro and B.~Schenke for discussions and collaborations on related work. This work was supported in part by the German Research Foundation (DFG). S.S.~and R.V.~are supported by US Department of Energy under DOE Contract No.~DE-AC02-98CH10886. K.B.~thanks HGS-HIRe for FAIR for support. S.S.~gratefully acknowledges a Goldhaber Distinguished Fellowship from Brookhaven Science Associates. R.V.~thanks Heidelberg University for hospitality and support as an Excellence Initiative Guest Professor, and the ExtreMe Matter Institute EMMI for support as an EMMI Visiting Professor.



\begin{thebibliography}{10}

\bibitem{Schmiedmayer:2013xsa}
  J.~Schmiedmayer and J.~Berges,
  Science {\bf 341} (2013) 6151,  1188.

\bibitem{Braaten:2004rn}
  E.~Braaten and H.~-W.~Hammer,
  Phys.\ Rept.\  {\bf 428} (2006) 259.
  
\bibitem{Berges:2008wm}
  J.~Berges, A.~Rothkopf and J.~Schmidt,
  Phys.\ Rev.\ Lett.\  {\bf 101} (2008) 041603.  
  
\bibitem{Berges:2008sr}
  J.~Berges and G.~Hoffmeister,
  Nucl.\ Phys.\ B {\bf 813} (2009) 383.  
  
\bibitem{Scheppach:2009wu}
  C.~Scheppach, J.~Berges and T.~Gasenzer,
  Phys.\ Rev.\ A {\bf 81} (2010) 033611.   

\bibitem{Nowak:2010tm}
  B.~Nowak, D.~Sexty and T.~Gasenzer,
  Phys.\ Rev.\ B {\bf 84} (2011) 020506(R).  

\bibitem{Berges:2012us}
  J.~Berges and D.~Sexty,
  Phys.\ Rev.\ Lett.\  {\bf 108} (2012) 161601.  
  
\bibitem{Nowak:2012gd}
  B.~Nowak, J.~Schole and T.~Gasenzer,
  arXiv:1206.3181v2 [cond-mat.quant-gas].  
  
\bibitem{Gelis:2010nm}
  F.~Gelis, E.~Iancu, J.~Jalilian-Marian and R.~Venugopalan,
  Ann.\ Rev.\ Nucl.\ Part.\ Sci.\  {\bf 60} (2010) 463.
  
\bibitem{Lappi:2006fp}
  T.~Lappi and L.~McLerran,
  Nucl.\ Phys.\ A {\bf 772} (2006) 200. 

\bibitem{Berges:2008mr}
  J.~Berges, S.~Scheffler and D.~Sexty,
  Phys.\ Lett.\ B {\bf 681} (2009) 362.

\bibitem{Schlichting:2012es}
  S.~Schlichting,
  Phys.\ Rev.\ D {\bf 86} (2012) 065008.
  
\bibitem{Kurkela:2012hp}
  A.~Kurkela and G.~D.~Moore,
  Phys.\ Rev.\ D {\bf 86} (2012) 056008.

\bibitem{Berges:2013eia}
  J.~Berges, K.~Boguslavski, S.~Schlichting and R.~Venugopalan,
  Phys.\ Rev.\ D {\bf 89} (2014) 074011; {\it ibid.} 114007.
  
\bibitem{Kurkela:2014tea}
  A.~Kurkela and E.~Lu,
  arXiv:1405.6318 [hep-ph].    
  
\bibitem{Endres:2012wg}
  M.~Endres {\it et al.},
  Nature {\bf 487} (2012) 454.

\bibitem{Micha:2004bv}
  R.~Micha and I.~I.~Tkachev,
  Phys.\ Rev.\ D {\bf 70} (2004) 043538.

\bibitem{Berges:2013lsa}
  J.~Berges, K.~Boguslavski, S.~Schlichting and R.~Venugopalan,
  JHEP {\bf 1405} (2014) 054.
    
\bibitem{waveturbulence}
  S.~Nazarenko, ``Wave Turbulence,'' Springer-Verlag (2011). 

\bibitem{Baier:2000sb}
  R.~Baier, A.~H.~Mueller, D.~Schiff and D.~T.~Son,
  Phys.\ Lett.\ B {\bf 502} (2001) 51.
  
\bibitem{BBSV_in-prep}
  J.~Berges, K.~Boguslavski, S.~Schlichting and R.~Venugopalan,
  in preparation  
  
\bibitem{Blaizot:2011xf}
  J.~-P.~Blaizot, F.~Gelis, J.~-F.~Liao, L.~McLerran and R.~Venugopalan,
  Nucl.\ Phys.\ A {\bf 873} (2012) 68.

\bibitem{Berloff}  
  N.~G.~Berloff, B.~V.~Svistunov, Phys.\ Rev.\ A {\bf 66} (2002) 013603.     
  
    
\end{thebibliography}
\end{document}